\documentclass[aps,prl,10pt,twocolumn,superscriptaddress,amsmath,amssymb]{revtex4-2}

\usepackage{graphicx,bm,braket,mathrsfs,orcidlink,pgfplots,xcolor,times}
\usepackage[normalem]{ulem}

\newcommand{\fkHz}{f_{_{F}}}
\newcommand{\nEOS}{$\sim5\times10^4$}

\newcommand{\ie}{i.e.,~}
\newcommand{\eg}{e.g.,~}

\begin{document}

\title{Multi-messenger and multi-band signal from first-order phase
  transitions in proto-neutron stars}

\author{\orcidlink{0000-0002-8669-4300}Christian Ecker}
\affiliation{Institut f\"ur Theoretische Physik, Goethe Universit\"at,
  Max-von-Laue-Str. 1, 60438 Frankfurt am Main, Germany}

\author{\orcidlink{0000-0001-6190-6947}Mauro Giliberti}
\affiliation{Institut f\"ur Theoretische Physik, Goethe Universit\"at,
  Max-von-Laue-Str. 1, 60438 Frankfurt am Main, Germany}

\author{\orcidlink{0000-0002-1330-7103}Luciano Rezzolla}
\affiliation{Institut f\"ur Theoretische Physik, Goethe Universit\"at,
  Max-von-Laue-Str. 1, 60438 Frankfurt am Main, Germany}

\affiliation{School of Mathematics, Trinity College, Dublin 2, Ireland}

\affiliation{Frankfurt Institute for Advanced Studies,
  Ruth-Moufang-Str. 1, 60438 Frankfurt am Main, Germany}

\begin{abstract}
  A first-order phase transition (PT) alters a compact star on two
  scales: globally, via growth of a quark-matter core, and
  microscopically, via collision of quark bubbles, driving a simultaneous
  gravitational-wave (GW) emission in the kHz and MHz bands. Using a
  constrained ensemble of model-agnostic equations of state, we follow
  the accretion-driven evolution of proto-neutron stars through the PT
  and compute the resulting multi-band GW signal. The correlations found
  between the kHz emission and the MHz burst could help constrain nuclear
  matter and the physics of the PT. A delayed neutrino burst should
  accompany the signal, with the delay set by the details of the PT, thus
  providing a prime multi-messenger source.
\end{abstract}

\maketitle

\noindent\textit{Introduction}---The composition of neutron-star cores is
one of the central open questions in nuclear astrophysics and
multimessenger astronomy. A first-order phase transition (PT) from
hadronic to quark matter would alter the star on two widely separated
lengthscales: globally, through the formation and growth of a
quark-matter core, and microscopically, if the conversion proceeds
through the nucleation, expansion, and collision of quark-matter
bubbles. These mechanisms can generate gravitational waves (GWs) in
distinct frequency bands. The rapid increase in neutron-star compactness
and associated structural readjustment induced by the PT can excite
global stellar modes in the kilohertz (kHz) range, while bubble
collisions and the resulting acoustic motion can generate GWs at
megahertz (MHz) frequencies. While PT-induced stellar collapse has long
been known to produce kHz oscillations~\cite{Abdikamalov2009b, Janka12,
  Sagert2011}, the MHz mechanism was proposed only recently for
binary-neutron-star mergers~\cite{Blas:2022xco} and subsequently
investigated in nascent neutron stars~\cite{Bleau:2026ala}. Similar
multi-band emission may therefore arise both in newly formed neutron
stars and in binary-neutron-star merger remnants~\cite{Baiotti2016,
  Radice2020b}.

While the kHz signal is already accessible to ground-based GW
interferometers, present detectors such as Advanced LIGO and Virgo have
limited sensitivity at these frequencies~\cite{Ganapathy2021}.
Third-generation (3G) detectors, such as the Einstein
Telescope~\cite{abac2025_etal} and Cosmic Explorer~\cite{Evans2021},
should substantially improve their reach, enabling frequent detections of
kHz GW signals at high signal-to-noise ratios. The MHz regime instead
lies beyond conventional interferometric detectors and requires dedicated
high-frequency concepts~\cite{Aggarwal:2025noe, Budker:2026uec}. This
experimental separation is mirrored by the theoretical description of the
two signals. The kHz component has been studied extensively, both
perturbatively and through numerical simulations, and is primarily
governed by the bulk equation of state (EOS), the resulting change in
stellar structure, and the stellar spin [see, e.g.,~\cite{Kokkotas99b,
    Baiotti2016, Radice2020b} for reviews]. The MHz component depends
additionally on poorly constrained microscopic properties of phase
conversion, most notably the surface tension, bubble dynamics, and
bubble-wall velocity, and involves scales that remain inaccessible to
global numerical-relativity simulations. Consequently, with only two
notable exceptions~\cite{Blas:2022xco, Bleau:2026ala}, the MHz signal
remains largely unexplored.

Previous studies of PT-induced GW emission have focused exclusively on
either the kHz~\cite{Abdikamalov2009b, Most:2018eaw, Bauswein:2018bma,
  Weih:2019xvw, Prakash:2021wpz, Espino2024c} or the MHz
regime~\cite{Blas:2022xco, Bleau:2026ala}, i.e., they have treated the
two frequency bands in isolation, obscuring their common origin and
preventing one from constraining the interpretation of the other. This
complementarity is particularly relevant for proto-neutron stars (PNSs),
where continued accretion after core bounce can trigger a delayed PT. The
resulting multimessenger (MM) sequence will comprise an initial kHz GW
signal and neutrino burst at bounce, followed by a second neutrino burst
together with a kHz and MHz GW emission if a PT occurs~\cite{Fischer2018,
  Sagert2011, Largani2024}. At the same time, increasingly precise
first-principles constraints from Chiral Effective Field Theory
(ChEFT)~\cite{Drischler:2021kxf} and perturbative QCD
(pQCD)~\cite{Gorda:2023mkk} combined with multimessenger observations
have begun to strongly constrain the cold neutron-star
EOS~\cite{Annala2019, Raaijmaers2021b, Ecker:2022b, Huth2022,
  Annala:2022, Somasundaram:2022, Komoltsev:2024lcr, Brandes:2024wpq,
  Rutherford:2024srk, Blomqvist2025, Gorda:2025aiu, Fernandez:2026hzs}.
These advances have been propagated to the kHz post-merger GW signal and
correlated with the EOS at the highest densities reached in neutron
stars~\cite{Ecker:2024b}. Their implications for the MHz GW emission,
however, remain unexplored.

In this Letter, we construct the PT-induced kHz and MHz components from
the same posterior-weighted, model-agnostic ensemble of cold,
beta-equilibrated EOSs constrained by ChEFT close to nuclear saturation
density \hbox{$n_{\rm sat}=0.16\,{\rm fm}^{-3}$}, and by pQCD far beyond
neutron-star core-densities. We select models supporting stable
quark-matter cores and, as a controlled first step, follow prescribed
trajectories in pressure and density at zero temperature. The metastable
hadronic branch determines the nucleation history and MHz spectrum, while
the corresponding pre- and post-transition configurations set the kHz
$F$-mode ringdown. The posterior weights of our Bayesian EOS analysis are
then inherited by the corresponding kHz and MHz signals, yielding a
posterior-informed multi-band prediction.

The two GW bands probe complementary physics: the kHz signal is governed
mainly by the bulk structural change of the star, whereas the MHz
spectrum is sensitive to microscopic PT dynamics and the rate at which
the PNS is driven through the metastable phase. A kHz measurement can
therefore reduce the EOS uncertainty in the interpretation of the MHz
signal, while a coincident MHz burst would probe bubble-mediated
conversion, including the surface tension and wall dynamics. Furthermore,
in a PNS, the accompanying neutrino emission provides an additional MM
trigger and timing information that can help identify the PT.

\begin{figure}
  \includegraphics[width=\columnwidth]{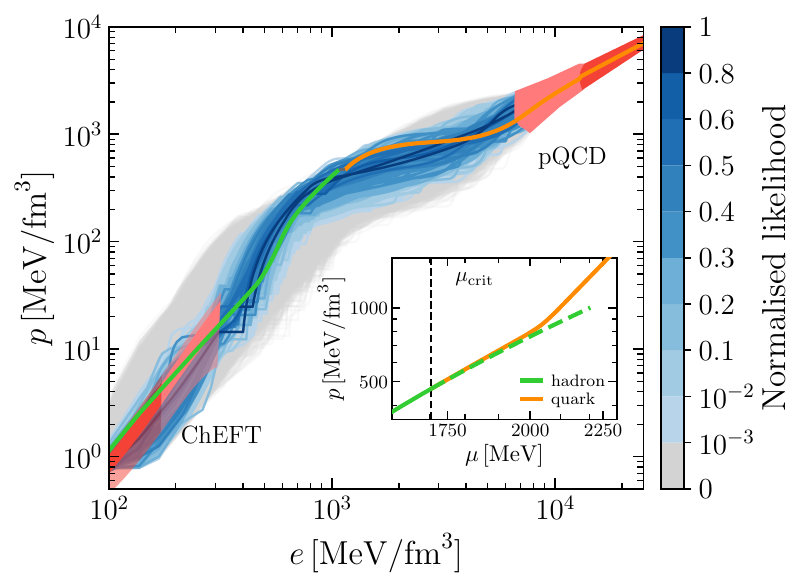}
  \caption{Model-agnostic cold EOS ensemble used in the analysis. Colours
    indicate the posterior likelihood from the imposed astrophysics and
    ChEFT constraints; the prior is shown in grey. Dark red shading marks
    the pQCD pressure constraints, while light red marks the
    better-converged sound-speed constraint, both entering the prior
    construction. For the ChEFT low and high transparency indicate one-
    and two-$\sigma$ uncertainties, respectively (see SM for details).}
\label{fig:fig_1}
\end{figure}

\noindent\textit{Agnostic EOSs and PT modelling}--- We use the
sound-speed parametrization~\cite{Annala2019,Altiparmak:2022} to
construct a model-agnostic ensemble of cold EOSs with a single
first-order PT between hadronic and quark matter, denoted by subscripts
$h$ and $q$, respectively~\cite{Blomqvist2025} (see Fig.~1).  At the
transition chemical potential $\mu_{\rm crit}$ the hadronic and quark
phases coexist in mechanical and chemical equilibrium, requiring
\hbox{$\mu_h=\mu_q=:\mu_{\rm crit}$}, and \hbox{$p_h(\mu_{\rm crit}) =
  p_q(\mu_{\rm crit})=:p_{\rm crit}$} (see inset in
Fig.~\ref{fig:fig_1}).  Because the transition is first order, the baryon
number and energy densities are discontinuous across the phase boundary,
jumping from $(n_h, e_h)$ in the hadronic phase to $(n_q, e_q)$ in the
quark phase. We here focus on PTs that preserve stellar stability at
their onset, which requires the Seidov condition~\cite{Seidov:1971} $e_q
< 3(e_h + p_{\rm crit})/2$. This choice avoids twin-star
branches~\cite{Alford2013, Christian2018, Montana2018, Haque2026} and
yields stars in which quark cores can grow continuously with increasing
central density. The selected sample contains \nEOS~EOSs with stable
quark-matter cores, whose thermodynamic and stellar properties are
summarised in the Supplemental Material (SM).

In the scenario we envision, the collapse of a massive stellar core forms
an initially hadronic PNS that undergoes a \textit{delayed} PT to quark
matter as continued accretion of infalling material raises its central
density (see, \eg~\cite{Janka12, Shankar2025}). To model this delayed
conversion, we extend the hadronic branch of the
Tolman--Oppenheimer--Volkoff solutions into the metastable region, $\mu
>\mu_{\rm crit}$, by extrapolating the last segment of the sound-speed
parameterisation preceding the PT by up to $500\,\mathrm{MeV}$ in
chemical potential, or terminating earlier if the sound speed reaches
either zero or the speed of light. To model the evolution into the
metastable region, we account for the accretion-driven structural changes
by evolving the stellar baryon mass $M_b$ from its initial value
$M_{b,0}$ at an accretion rate $\dot{M}_b$ (see SM).  Note that baryon
mass rather than gravitational mass $M$ is the natural evolution variable
that is conserved during the PT, whereas $M$ changes with the stellar
binding energy \footnote{Using the gravitational mass as the accreted
quantity is therefore not equivalent to evolving $M_b$, since generally
$dM/dM_b\neq1$. This modifies the mapping from time to central chemical
potential and hence $t_{\rm crit}$, $t_{\rm nuc}$, and the overpressure
$\Delta p(t)$, propagating to $N_{\rm bbl}$ and $R_{\rm bbl}$. The effect
can be amplified because the thin-wall action scales as
$S\propto\sigma^4/[\Delta p(t)]^3$ and enters exponentially in the
nucleation rate, making it sensitive to even modest changes in $\Delta
p(t)$.}.  Since the accretion history depends on the details of the
collapse, we model it randomly assuming a uniform prior distribution
$\dot{M}_{b} \in [0.2, 1.0] \,M_{\odot}/{\rm s}$~\cite{Shankar2025}. In
this way, we can map the evolution of $M_b(t)$ onto an evolution of the
central chemical potential $\mu_c(t)$ along the metastable sequence. The
PT can take place after the critical time $t_{\rm crit}$ at which the
overpressure \hbox{$\Delta p(t) := p_q[\mu_c(t)]-p_h[\mu_c(t)]$} is
zero. Since the quark-bubble nucleation process is exponentially
suppressed at $t_{\rm crit}$, a sufficient amount of overpressure is
needed to favour bubble nucleation and trigger the PT at $t_{\rm nuc} >
t_{\rm crit}$ (see SM). Since in our analysis thermal effects are
neglected, the conversion is modelled as zero-temperature quantum
nucleation in hadronic matter~\cite{Iida:1997ay}, with the nucleation
action being that of a thin-wall bubble~\cite{Coleman:1977py}.
More specifically, we consider a bubble-nucleation rate per
unit volume given by~\cite{Blas:2022xco}
\begin{equation}
 \Gamma(t)=\mathcal{S}_{\rm nuc}^4 \exp{\left[-S(t)\right]} = 
 \mathcal{S}_{\rm nuc}^4 \exp
 {\left[ - \frac{27 \pi^2 \sigma^4}{2(\Delta p(t))^3}\right]}\,,
 \label{eq:rate}
\end{equation}
where $S(t)$ is the thin-wall action, $\sigma$ is the interface tension,
and $\mathcal{S}_{\rm nuc} = 200\,\mathrm{MeV}$ is a constant reference scale. 
This approximation follows previous treatments~\cite{Blas:2022xco, Bleau:2026ala} 
and provides a simple description of the scenario investigated here.

An important but poorly constrained quantity in quark-bubble dynamics is
the wall velocity $v_w$, which governs their expansion in the hadronic
medium. This velocity, which can either be supersonic (detonation) or
subsonic (deflagration)~\cite{Rezzolla_book:2013}, is here treated as a
free parameter and determines the fraction of supercritical matter which
remains in the metastable hadronic phase at $t>t_{\rm
  crit}$~\cite{Enqvist1992}, \ie
\begin{equation}
 q(t) := \exp\!\left[-\frac{4\pi}{3}v_w^3 \int_{t_{\rm crit}}^{t} 
   \, \Gamma(t')\,  (t-t')^3 \, dt'\right] \,.
 \label{eq:q(t)}
\end{equation}
In practice, we find that for the conditions leading to a viable PT
scenario, the wall velocity is typically subsonic. The asymptotic bubble
density and mean separation at collision are then defined as $n_{\rm bbl}
:= \int_{t_{\rm crit}}^{\infty}d t\,q(t) \, \Gamma(t)$, and $R_{\rm bbl}
:= n^{-1/3}_{\rm bbl}$, respectively. Furthermore, the number of bubbles
is given by $N_{\rm bbl} := (R_q/R_{\rm bbl})^3$, where $R_q$ is the
radius of the quark core at $t_{\rm nuc}$.

\begin{figure}
  \includegraphics[width=0.48\textwidth]{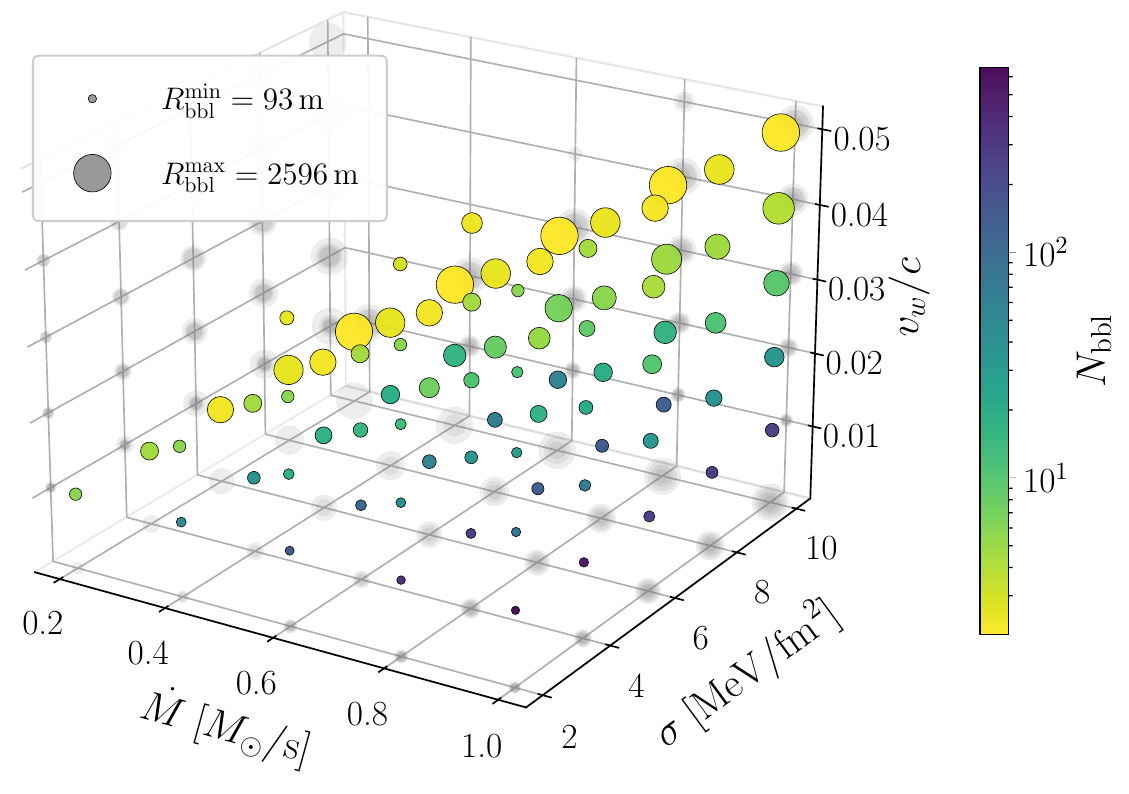}
  \caption{Bubble statistics for the representative EOS highlighted in
    Fig.~\ref{fig:fig_1}, under variations of $(\sigma, v_w,
    \dot{M}_b)$. The colour of the marker denotes the number of bubbles
    in the converted core and its size the mean bubble radius at
    collision; the minimum and maximum values for the representative EOS
    are indicated in the legend.}
\label{fig:bubbles}
\end{figure}

In summary, once a specific EOS has been selected, our parameterised
representation of the PTs depends on three independent degrees of
freedom: the mass-accretion rate $\dot{M}_b$, the bubble surface tension
$\sigma$, and the wall velocity $v_w$. In turn, these quantities set the
number of bubbles produced $N_{\rm bbl}$ and their average size at
collision $R_{\rm bbl}$. Figure~\ref{fig:bubbles} provides a unified
representation of these five quantities for the representative EOS
highlighted in Fig.~\ref{fig:fig_1}, with colours indicating $N_{\rm
  bbl}$ and marker size encoding $R_{\rm bbl}$. Overall,
Fig.~\ref{fig:bubbles} shows that the bubble properties do not correlate
in a simple way with either the microscopic PT parameters ($\sigma, v_w$)
or the macroscopic properties (EOS, $\dot{M}_b$). This means that a
single EOS or microscopic PT model cannot capture the resulting spread in
bubble properties. Instead, jointly sampling the EOS ensemble and the
parameters ($\sigma, v_w, \dot{M}_b$) allows us to identify robust
features of the GW signal across these uncertainties.

\noindent\textit{MHz and kHz GW emission}---To model the GW emission, we
treat separately the low- and high-frequency signals, recalling that the
kHz signal will be emitted twice: first when the PNS is formed (and which
we do not model here) and then at the time of the PT. To estimate the
latter, for each PT we match the metastable and hybrid configurations at
fixed baryon mass and angular momentum, while scanning over the different
masses at nucleation across the ensemble. Quasi-universal relations for
hybrid stars can be used to estimate the typical $F$-mode frequencies and
their damping times in terms of the tidal deformability $\Lambda_q$ of
the hybrid star~\cite{Zhao:2022tcw}. In turn, the knowledge of these
frequencies and of the difference in the hadron and hybrid stellar
quadrupole moments can be used to compute the corresponding
\textit{characteristic frequency} and \textit{strain} of the kHz signal,
\ie $f_{\rm ch, k}$ and $h_{\rm ch, k}$ (see SM).

On the other hand, to model the MHz GW signal, we retain only
microphysical realisations that lead to a bubble number $N_{\rm bbl} \geq
2$ as this is the minimum needed for bubbles to generate collisions and
hence a GW emission. Once $N_{\rm bbl}$ and $R_{\rm bbl}$ are computed,
the full GW spectrum from the PT is evaluated with a simple
sound-shell-model-inspired formula~\cite{Hindmarsh:2019phv,
  Hindmarsh2021} whose scale is set by $R_{\rm bbl}$
and whose amplitude depends on the kinetic-energy density of the
quark matter fluid $\rho_{\rm kin}$, which we set to be a fixed fraction
of $\mathcal{S}_{\rm nuc}^4$~\cite{Blas:2022xco}, and on the size of
the converted quark core $R_q$ (see SM).

\begin{figure}
  \centering
  \includegraphics[width=0.48\textwidth]{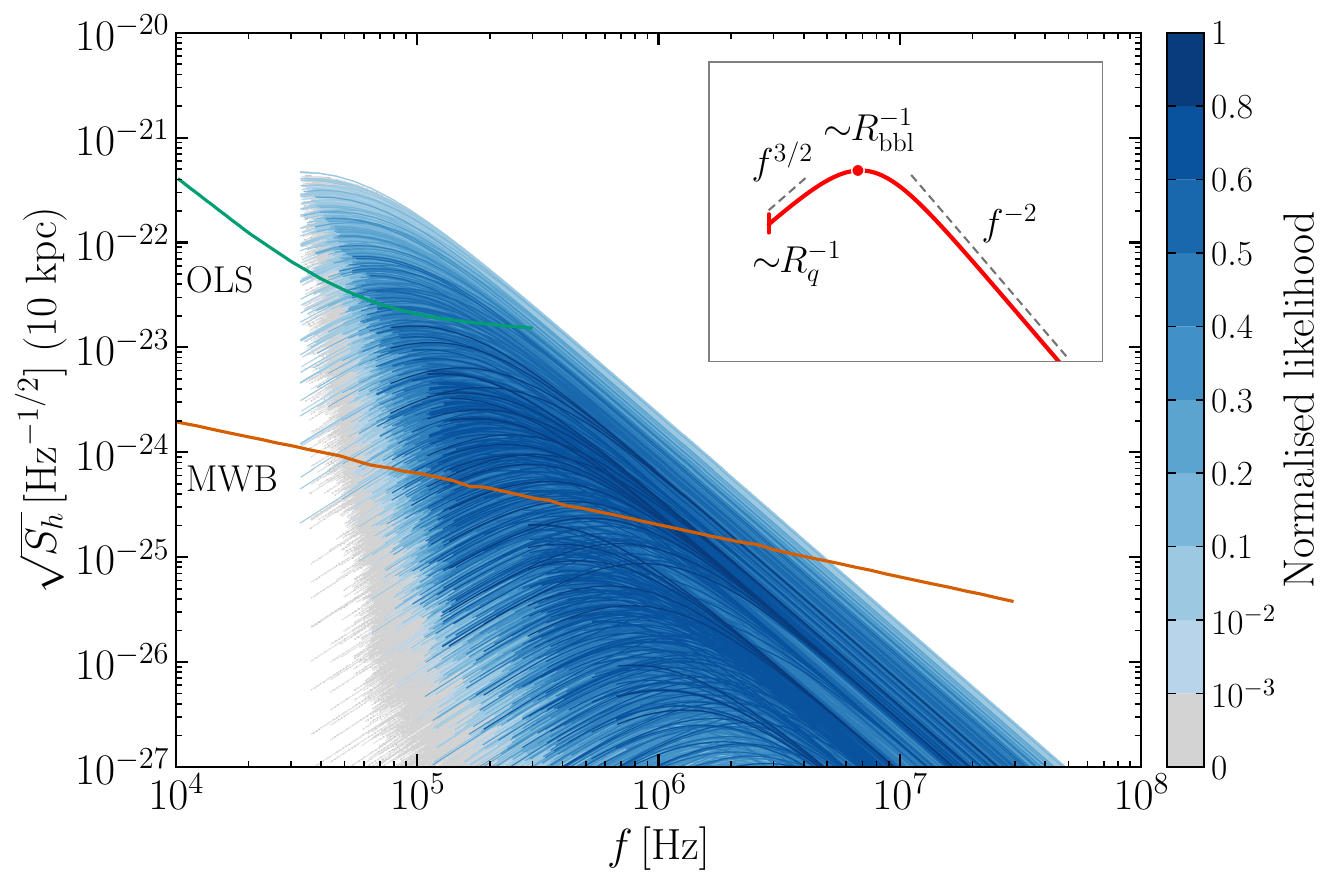}
  \caption{MHz strain frequency spectra for the same EOS ensemble as in
    Fig.~\ref{fig:multi-band}, with the same colour-coded likelihood; 
    also shown are the
    sensitivity curves of high-frequency detectors. The inset illustrates
    the properties of a single spectrum, highlighting the main features.}
  \label{fig:straincurves}
\end{figure}

Figure~\ref{fig:straincurves} shows the spectrum of the MHz GW signal for
\nEOS~EOSs and their corresponding normalised likelihood. Also, in order
to highlight the signal's features, we show in the inset a representative
spectrum for a single EOS and choice of $\sigma, v_w, \dot{M}_b$. In
particular, the low-frequency cutoff is imposed as the inverse of the
size of the quark core $R^{-1}_q$, while the frequency at maximum strain
$f_{\rm max, M}$ is set by the inverse of the mean bubble separation at collision
$R^{-1}_{\rm bbl}$. Similarly, the broken power law reflects the complex
contribution from the expansion and collision of the
bubbles~\cite{Hindmarsh:2019phv}, where the low-frequency slope is fixed
by causality, while the high-frequency one is the decay measured in fluid
lattice simulations~\cite{Hindmarsh:2017Shape}. Using the full
information of the GW signal, we can compute the characteristic frequency
and strain in the MHz range, \ie $f_{\rm ch, M}$ and $h_{\rm ch, M}$,
which are useful quantities to compare with the corresponding quantities
in the kHz band. Interestingly, a tight correlation exists between
$f_{\rm ch, M}$ and $f_{\rm max, M}$, so that a measurement of the former
provides immediate information on the typical bubble size at collision
$R_{\rm bbl}$ (see Fig.~\ref{fig:fig_1EM} in End
Matter).

\begin{figure*}[htb]
  \centering
  \includegraphics[height=0.22\linewidth]{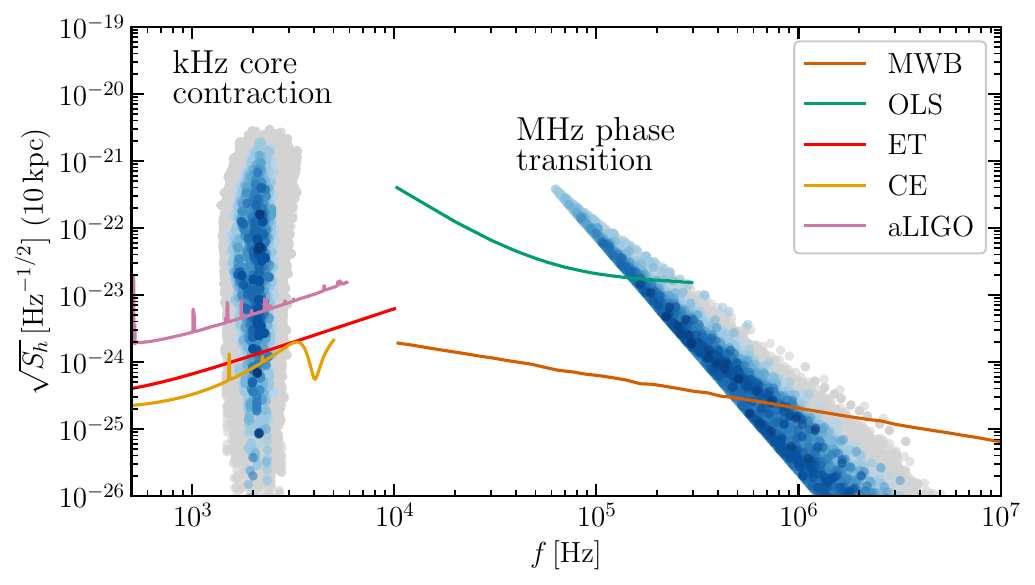}
  \includegraphics[height=0.22\linewidth]{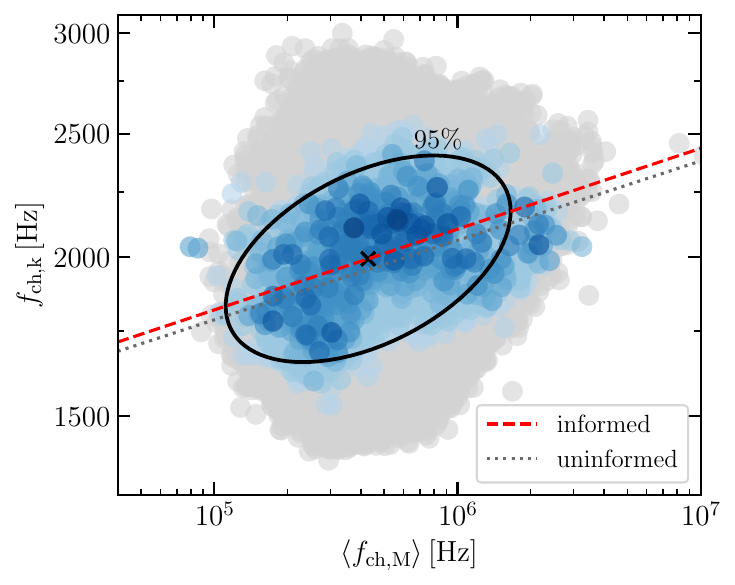}
  \includegraphics[height=0.22\linewidth]{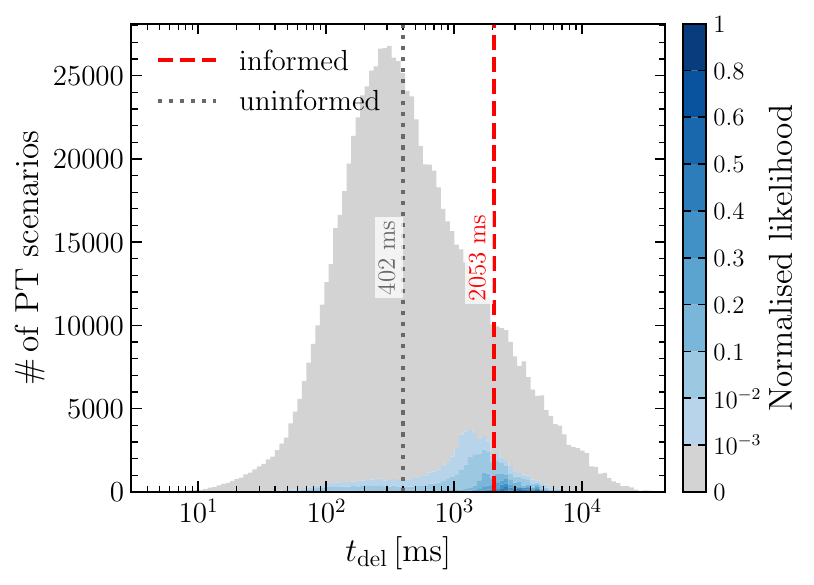}
  \caption{\textit{Left:} Characteristic strains and frequencies in their
    kHz and MHz components with the colormap reporting the corresponding
    normalised likelihood. Shown with solid lines are the
    sensitivities~\cite{Aggarwal:2025noe, Muia:2025HFGWPlotter} for
    Advanced LIGO (aLIGO)~\cite{Effler2025}, Einstein Telescope
    (ET)~\cite{Hild2011}, Cosmic Explorer (CE)~\cite{Srivastava2022},
    resonant magnetic Weber bars (MWB)~\cite{Domcke2025}, and optically
    levitated sensors (OLS)~\cite{Aggarwal2022} (the sources are assumed
    to be at a distance of $10\,\mathrm{kpc}$). \textit{Middle:}
    Correlation between the characteristic kHz frequency $f_{\rm ch,k}$
    and the mean MHz frequency $\langle f_{\rm ch,M}\rangle$. The black
    ellipse encloses the $95\%$ credibility region with the cross marking
    its mean, while the dashed red and dotted grey lines are power-law
    fits to the likelihood-informed and the uninformed distributions,
    respectively. \textit{Right:} Distribution of the neutrino-burst
    delay time $t_{\rm del} \simeq t_{\rm nuc}$, across the various PT
    scenarios. The dotted grey and dashed red lines highlight the median
    of the full or of the likelihood-weighted distributions,
    respectively.}
  \label{fig:multi-band}
\end{figure*}

Figure~\ref{fig:multi-band} summarises the multi-band and MM predictions
of our analysis by reporting in the left panel the characteristic
frequencies and strains in their low- and high-frequency components,
along with the corresponding normalised likelihood (the sources are
assumed to be at a distance of $10\,\mathrm{kpc}$). Also shown are the
sensitivities for representative interferometric detectors, such as
Advanced LIGO (aLIGO), the Einstein Telescope (ET), and Cosmic Explorer
(CE), as well as resonant magnetic Weber bars (MWB), and optically
levitated sensors (OLS)~\cite{Aggarwal:2025noe,
  Muia:2025HFGWPlotter}. The wealth of information in
Fig.~\ref{fig:multi-band} calls for a number of considerations. First,
although both bands are constructed from the same pre- and
post-transition stellar models, each kHz signal leads to a family of MHz
emissions, obtained when varying the degrees of freedom of the GW model:
$\sigma,v_w,\dot{M}_b$ (see Fig.~\ref{fig:fig_2EM} in End
Matter). Second, the likelihood of a kHz signal is correlated with that
of the MHz counterpart, that is, a high-likelihood kHz signal will lead
to a family of MHz signals having high-likelihood characteristic
frequencies and strains. Third, while the estimated high-frequency
signals are generally weaker than their low-frequency counterparts, they
are in principle detectable by the planned resonant MWB and OLS
experiments. Fourth, the variance in frequency of the kHz and MHz signals
is significantly different, with the former varying by less than a factor
of two and the latter by almost two orders of magnitude. This is because
while the kHz signal only weakly depends on the global properties of the
stellar models, which vary by a factor of two at most, the MHz component
follows the bubble dynamics, which depends exponentially on the interface
tension and on how rapidly the star is driven through the metastable
branch (see End Matter).

Despite the considerably different variance of the characteristic
frequencies in the two bands, a correlation emerges between $f_{\rm ch,
  k}$ and $\langle f_{\rm ch, M} \rangle$, where the latter is obtained
by averaging for each EOS the characteristic MHz frequency over the
sampled phase-conversion parameters. As shown in the middle panel of
Fig.~\ref{fig:multi-band}, the ChEFT and astrophysical constraints
substantially reduce the high-likelihood region in the $(\langle f_{\rm
  ch,M}\rangle,f_{\rm ch,k})$ plane. The correlation in the distribution
is approximately described by
\begin{equation}
 \log_{10}\!\left(\frac{f_{\rm ch,k}}{\rm Hz}\right)
 \simeq k_1 +
 k_2\log_{10}\!\left(\frac{\langle f_{\rm ch,M}\rangle}{\rm Hz}\right),
\end{equation}
with $k_1 \simeq 2.93 (2.94)$ and $k_2 \simeq 0.062 (0.064)$ for the
posterior (prior), while the black ellipse indicates the corresponding
$95\%$ posterior region. This correlation provides a link between the two
frequency bands and illustrates how the EOS information encoded in the
kHz signal restricts the range of MHz frequencies associated with the
same transition.

Finally, we comment briefly on the MM aspects of the scenario we have
investigated, recalling that, in addition to GWs, it will offer two
distinct bursts of neutrinos. The first one will be emitted at PNS
formation~\cite{Janka12}, while the second one will follow the core
contraction at the PT~\cite{Sagert2011}. This second neutrino signal has
been explored only in a handful of numerical simulations (see,
\eg~\cite{Fischer2018, Sagert2011, Largani2024}) and its properties
clearly depend on the largely unknown space of parameters that regulates
the PT. That said, a first estimate of the interval separating the two
neutrino bursts, which is much larger than their duration, can be
obtained in terms of the time needed for the PNS to go from its initial
baryon mass to that needed to trigger the PT. In our model, this delay
time is $t_{\rm del} \simeq t_{\rm nuc}$ and obviously depends on
$\sigma, v_w$, and $\dot{M}_b$. The right panel of
Fig.~\ref{fig:multi-band} reports the distribution of delay times, which
is broad but with a clear peak with median value $t_{\rm del} \sim 400
\,{\rm ms}$ that moves towards $\sim 2.1 \,{\rm s}$ when considering the
posterior distribution of likelihood-weighted EOS configurations. To the
best of our knowledge, this is the first time that such a delay time has
been estimated, which can increase the detection probability
significantly and thus the opportunity to extract MM information on the
details of the PT.

\noindent\textit{Conclusion}---We have constructed an agnostic
description of the multi-band GW emission from a PNS undergoing a
first-order PT to quark matter. In particular, using an ensemble of
parameterised cold EOSs conditioned on the theoretical and observational
constraints considered here, we have carried out a statistical study of
the occurrence of a PT when the stellar models follow specific
zero-temperature trajectories in the pressure-energy density $(p, e)$
plane. This has allowed us to map the thermodynamic history of an EOS to
a dynamical history of the stellar models and, in turn, to determine the
time of the PT and the properties of the multi-band GW emission.  The
latter has been modelled differently for the kHz region -- where it
depends on the macroscopical stellar properties ($M, R, \Lambda, Q$) --
and in the MHz region -- where it is influenced by the microscopic
features of the PT ($\sigma, v_w$) and by the evolution of the PNS
($\dot{M}_b$).

The ``big picture'' of the likelihood-weighted GW emission obtained in
this way has revealed the importance of a multi-band detection in which
the kHz signal probes the bulk structural change in the compact star,
whereas the acoustic MHz burst probes the bubble nucleation and wall
dynamics. Their joint evaluation provides therefore distinct information
that is otherwise degenerate when only one band is detected.

We have also argued that the multi-band GW emission will be accompanied
by a coincident second burst of neutrinos with a characteristic delay
time $t_{\rm del} \sim 2 \, {\rm s}$. The MM nature of the
signal makes the PT from a PNS a particularly exciting source for 3G GW
detectors such as ET or CE and should act as an additional motivation for
the development of experimental techniques and detecting facilities able
to measure GWs in the MHz range.

While it presents a first unified proof-of-principle description of the
multi-band and MM emission from a PT in a PNS, our work is also
minimalistic in various aspects. First, we have ignored for simplicity
the temperature dependence in the EOS, which is instead relevant between
PNS formation and the triggering of the PT. Second, we have employed a
rather crude modelling of both the kHz and MHz emission; although we
expect the estimates made in terms of the characteristic strains and
frequencies to be qualitatively correct, precise quantitative predictions
may change with improved modelling. Third, while we have performed an
initial exploration of the correlations between the properties of the two
GW signals and between the MHz signal and the microphysical properties of
the PT, our analysis has been mostly phenomenological. Finally, while it
has emerged that the MHz signal is in principle detectable by novel GW
detectors such as OLS or MWB, our analysis has not been focused on
determining the space of parameters of the PT that are optimally detected
by these instruments.  All of these considerations call for exciting
improvements in our theoretical modelling that are beyond the scope of
this work but will find space in future research.

\medskip
\paragraph{Acknowledgements.} 
We thank Jorge Casalderrey-Solana, Mark Hindmarsh, Joachim Kopp, 
David Mateos, and Mikel Sanchez-Garitaonandia for useful
discussions. CE, MG and LR acknowledge support by the DFG through the
CRC-TR 211 ``Strong-interaction matter under extreme conditions'' --
project number 315477589 -- TRR 211. LR acknowledges support from the ERC
Advanced Grant ``JETSET: Launching, propagation and emission of
relativistic jets from binary mergers and across mass scales'' (grant
no. 884631). LR also acknowledges the Walter Greiner Gesellschaft zur
F\"orderung der physikalischen Grundlagenforschung e.V. through the Carl
W. Fueck Laureatus Chair.

\medskip
\paragraph{Data availability.} 

The \texttt{EOSsampler} code used to construct the EOS ensemble is
publicly available at Ref.~\cite{EOSsampler}; the \texttt{NSFOPT.jl} code
used to compute the kHz and MHz signals is publicly available at
Ref.~\cite{NSFOPT2026}. The data to produce the figures is available from
the authors upon reasonable request.
\bibliography{aeireferences}


\clearpage
\onecolumngrid
\vspace*{\fill}
\begin{center}
    \textbf{\large End Matter}
\end{center}
\vspace*{\fill}
\twocolumngrid
\appendix

\subsection*{Spectral correlations in the GW signals}

\begin{figure}[h!]
  \includegraphics[height=0.58\linewidth]{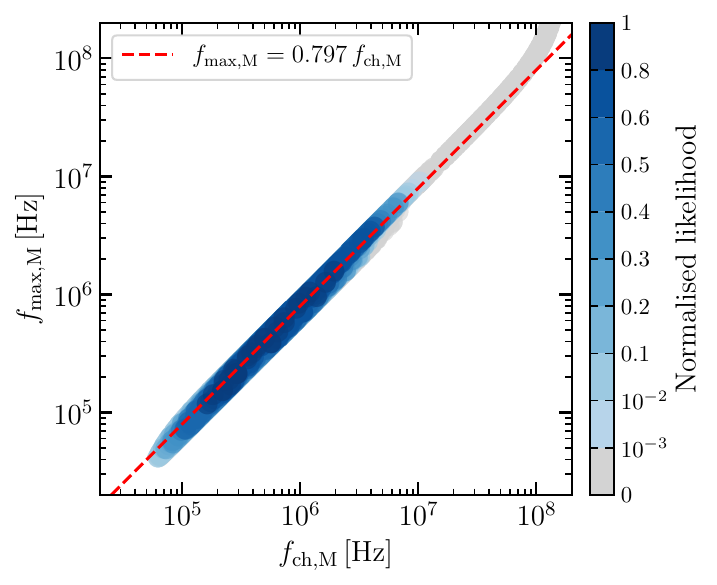}
    \caption{Likelihood-weighted correlation between the characteristic
      frequency $f_{\rm ch, M}$ and the frequency of maximum strain
      $f_{\rm max, M}$ of the MHz signal, for all considered EOSs
      and nucleation parameters.}
  \label{fig:fig_1EM}
\end{figure}

A particularly important result of our analysis lies in the spectral
correlations between the characteristic frequency in the MHz range
$f_{\rm ch, M}$ and the frequency of the maximum spectral density of the
full MHz signal $f_{\rm max, M}$. This is shown in
Fig.~\ref{fig:fig_1EM}, which highlights the tight correlation between
the two quantities with $f_{\rm max, M} \simeq 0.797 f_{\rm ch, M}$, and
reveals an important consequence: whatever the underlying microphysics of
the PT, and since $f_{\rm ch, M} \sim R^{-1}_{\rm bbl}$, measuring a
characteristic frequency of the MHz signal immediately provides direct
information on the typical bubble size at collision.

\begin{figure}[h!]
  \includegraphics[height=0.58\linewidth]{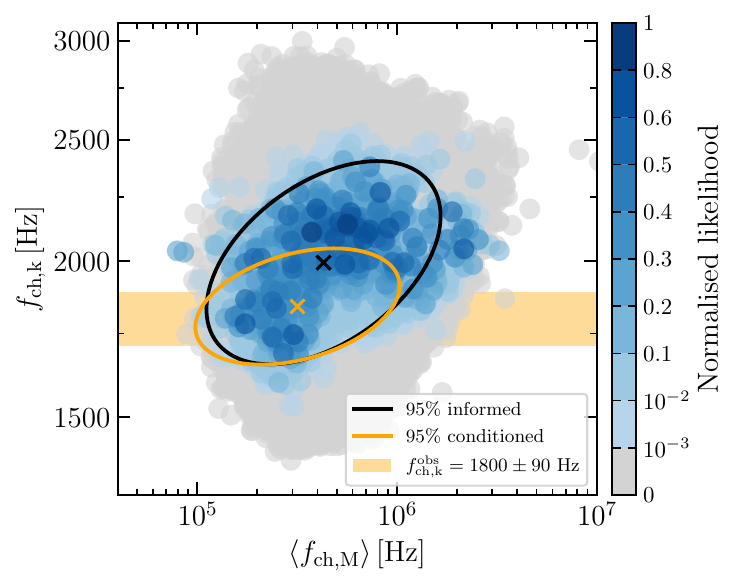}
    \caption{Change in the 95\% credible interval for the correlation
      between kHz and MHz characteristic frequencies after a conditioning
      mock measurement of $f_{\rm ch,k}^{\rm obs}$ with 5\% uncertainty,
      identified by the orange band; the new $95\%$ credibility ellipse
      is also marked in orange.}
  \label{fig:fig_4EM}
\end{figure}

To quantify the complementarity already discussed in
Fig.~\ref{fig:multi-band} (middle), we consider a mock measurement of the
kHz characteristic frequency, $f_{\rm ch, k}^{\rm obs}$, with relative
uncertainty of $5\%$, visible in Fig.~\ref{fig:fig_4EM}. We re-weight the
EOS ensemble with a Gaussian likelihood centred on $f_{\rm ch, k}^{\rm
  obs}$ and propagate the resulting weights to the MHz prediction. This
reduces the $95\%$ credible interval of $\langle f_{\rm ch,M}\rangle$
from $[1.81\times10^5, 1.54\times10^6]$ to $[1.66\times10^5,
  1.17\times10^6]$ Hz, corresponding to a reduction of $8.7\%$ in the
width of the interval in $\log_{10} f$. This means, even without
detecting the MHz component, a kHz measurement already restricts the
range of high-frequency signals compatible with the same PT. Overall,
these results underline the enormous advantages of multi-band GW
observations as the signals in the two bands constrain different aspects
of the same PT. In particular, the kHz signal constrains the pre- and
post-PT stellar properties and thereby reduces the EOS uncertainty
entering the MHz prediction.

\begin{figure}[h!]
  \includegraphics[height=0.58\linewidth]{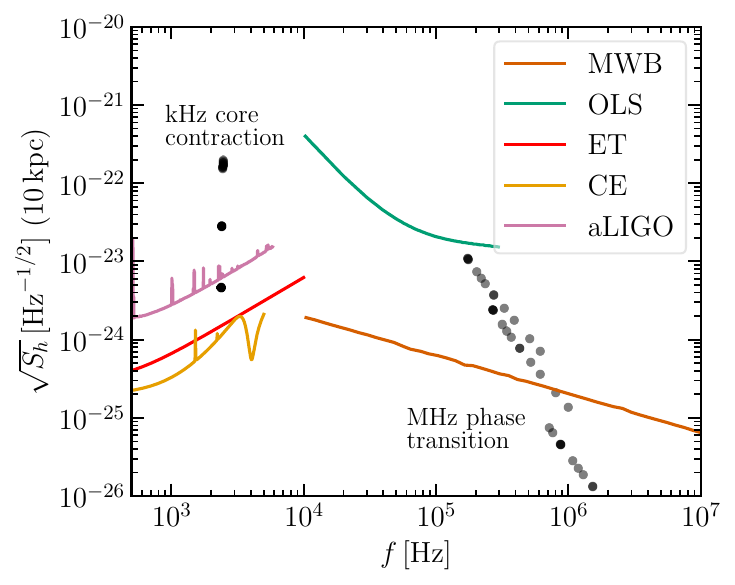}
  \caption{Characteristic strains and frequencies in the kHz and MHz
    bands, as shown in Fig.~\ref{fig:multi-band}, for the single
    representative EOS also shown in Figures~\ref{fig:fig_1}
    and~\ref{fig:bubbles}. Different markers are obtained by varying the
    PT parameters $\sigma$ (for both signals) and $v_w, \dot{M}_b$ (for
    the MHz signal).}
  \label{fig:fig_2EM}
\end{figure}

Finally, while Fig.~\ref{fig:multi-band} provides a statistical view of
the likelihood of the GW signal when considering an ensemble of \nEOS~
EOSs, it is interesting to analyse the correlations in the two bands in
response to a single EOS. This is shown in Fig.~\ref{fig:fig_2EM}, which
reports the characteristic strains and frequencies in the two bands for a
representative EOS. Starting from the low-frequency signals produced by
the PT-induced contraction of the PNS core, we note that the
characteristic frequency $f_{\rm ch,k}$ is barely affected by the
different stellar configurations leading to the PTs.
Indeed, both the frequency $f_{\rm ch,k}$ and the strain $h_{\rm ch,k}$
depend on the stellar mass at nucleation $M_{\rm nuc}$, which varies
mostly in response to differences in the surface tension (larger values
of $\sigma$ lead to larger overpressures, longer nucleation times and
hence larger masses at nucleation; see SM). However, the mass dependency
of the fundamental mode frequency is mostly absorbed into its dependency
on the tidal deformability $\Lambda_q$ of the star after the PT. On the
other hand, the kHz emission strain depends on the mass through the
quadrupole difference (see SM), which scales with the difference between
the two tidal deformabilities for the star pre- and post-conversion,
which is zero at criticality, resulting in a much stronger dependence on
$M_{\rm nuc}$. At the same time, the behaviour of the high-frequency GW
signal is significantly different since it depends in a highly nonlinear
manner also on the degrees of freedom of the bubble nucleation and
expansion model, namely, $\sigma$ and $v_w$. Hence, the characteristic
quantities span a range that is more than two orders in amplitude and
almost one in frequency.


\clearpage
\onecolumngrid
\setcounter{equation}{0}
\renewcommand{\theequation}{S\arabic{equation}}
\setcounter{figure}{0}
\renewcommand{\thefigure}{S\arabic{figure}}

\begin{center}
    \textbf{\large Supplemental Material}
\end{center}

\renewcommand{\theHfigure}{S\arabic{figure}}

\section{EOS ensemble construction and posterior weights}
\label{sec:S1}

In what follows, we briefly review the strategy followed to construct the
agnostic EOS ensemble and the compression histories needed for the GW
emission. The cold, beta-equilibrated EOS ensemble is constructed in
three different intervals in the matter number density, taking the nuclear
saturation density $n_{\rm sat}=0.16\,{\rm fm}^{-3}$ as a useful dividing
reference. In particular, below $n_{\rm match}=0.5\,n_{\rm sat}$, we use
a cold beta-equilibrium slice of the Steiner-Fischer-Hempel (SFHo)
nuclear EOS model ~\cite{Moller:1996uf,Steiner2013} from the CompOSE
database~\cite{Typel2015,CompOSECoreTeam:2022ddl}. At higher densities,
the squared sound speed is sampled as a piecewise-linear function of the
baryon chemical potential~\cite{Annala2019, Altiparmak:2022}
\begin{equation}
 \label{eq:S_cs}
 c_s^2(\mu) =
 \frac{(\mu_{i+1}-\mu)c_{s,i}^2+(\mu-\mu_i)c_{s,i+1}^2}{\mu_{i+1}-\mu_i}\,,
 \qquad \mu_i\leq\mu\leq\mu_{i+1}\,.
\end{equation}
Thermodynamic consistency determines the number density and pressure via
the relations
\begin{align}
  n(\mu)&=n_{\rm match}\exp\!\left[\int_{\mu_{\rm
        match}}^{\mu}\frac{d\mu'}{\mu'c_s^2(\mu')}\right]\,,
 \label{eq:S_nmu}\\
 p(\mu)&=p_{\rm match}+\int_{\mu_{\rm match}}^{\mu}d\mu'\,n(\mu')\,.
 \label{eq:S_pmu}
\end{align}
As in~\cite{Blomqvist2025}, we implement the low-density nuclear-theory
constraint using the conservative next-to-next-to-leading-order (N2LO)
ChEFT uncertainty band of Ref.~\cite{Drischler:2020fvz}, integrating the
corresponding pressure likelihood up to $1.2\,n_{\rm sat}$. Differently
from the approach in Ref.~\cite{Altiparmak:2022} but along the lines of
what was done in Ref.~\cite{Blomqvist2025}, for each EOS we introduce a
first-order PT at $\mu=\mu_{\rm crit},\,p=p_{\rm crit}$ in terms of a
discontinuity $\Delta n$ at fixed pressure and chemical potential,
separating the hadronic and quark-matter branches, that we denote with
the subscripts $h$ and $q$, respectively:
\begin{equation}
 p_q(\mu_{\rm crit}) = p_h(\mu_{\rm crit}) := p_{\rm crit}\,, \qquad
 n_q(\mu_{\rm crit}) = n_h(\mu_{\rm crit})+\Delta n\,,\qquad
 e_q(\mu_{\rm crit}) = e_h(\mu_{\rm crit})+\mu_{\rm crit}\Delta n\,.
 \label{eq:S_Maxwell}
\end{equation}
At high densities, we impose the pQCD constraints at $\mu_{\rm
  pQCD}=2.6\,\mathrm{GeV}$ following~\cite{Blomqvist2025}, including the
scale-averaging prescription of~\cite{Gorda:2023usm} and the additional
speed-of-sound information at $\mu=2.25$ and $2.43\,\mathrm{GeV}$
from~\cite{Komoltsev:2023zor}.

\begin{figure}[htb]
  \includegraphics[width=\textwidth]{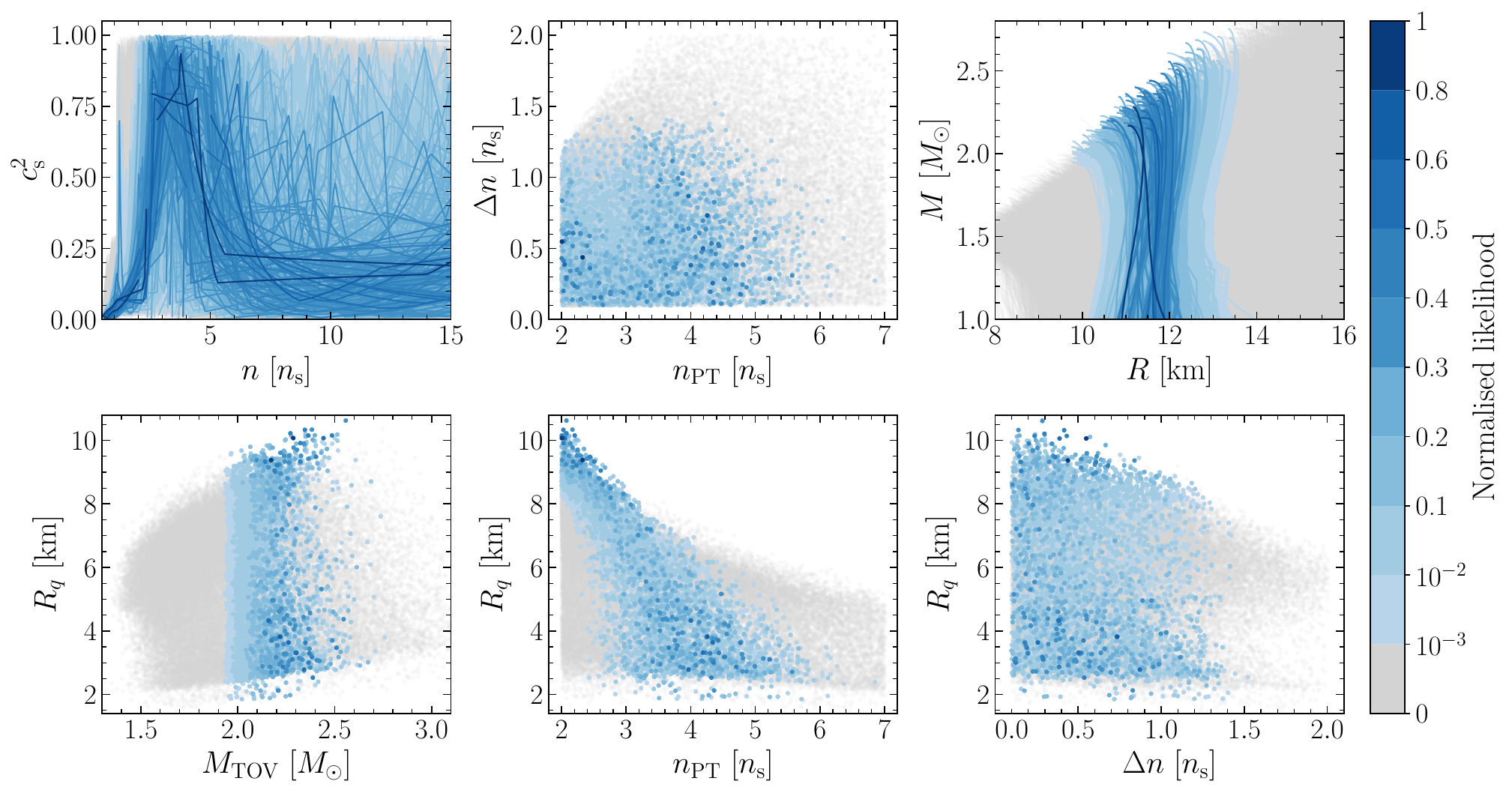}
  \caption{Posterior-weighted family of \nEOS~ zero-temperature EOS
    models that support stable neutron-star solutions with quark-matter
    cores, and sustain enough overpressure to nucleate. The panels show
    the speed of sound squared, PT parameters, mass--radius curves, and
    correlations of the maximum quark-core radius with ${M_{_{\rm
          TOV}}}$, transition density, and density discontinuity. Since
    EOSs with a short metastable branch do not reach the nucleation
    overpressure, low values of $R_{\rm q}$ are disfavoured.}
\label{fig:QMcoreEOSs}
\end{figure}

In addition, each EOS is conditioned with astrophysical observations
following the same approach and selection of observational data as
in~\cite{Blomqvist2025}. The mass likelihood ${\cal L}_{M}$ includes
radio measurements of the massive pulsar PSR
J1614$-$2230~\cite{Arzoumanian2018,Fonseca2016,Demorest2010}, while
${\cal L}_{\rm GW}$ incorporates the tidal-deformability constraint from
the binary-neutron-star merger GW170817~\cite{LIGOScientific:2018hze}.
The X-ray likelihood ${\cal L}_{X}$ combines mass--radius inferences from
the NICER pulsars PSR J0030+0451, PSR J0740+6620, PSR J0437$-$4715, and
PSR J0614$-$3329 \cite{MCMiller2019b, Riley_2019, Fonseca2021,
  Miller2021, Riley2021, Choudhury:2024xbk, Mauviard:2025dmd}, together
with constraints from quiescent low-mass X-ray binaries and X-ray
bursters \cite{Shaw:2018wxh, Steiner2017, Nattila2017, Nattila2016}. We
refer to Ref.~\cite{Annala2023} for details of the likelihood
construction. Each EOS $\alpha$ therefore carries the normalised
posterior weight
\begin{equation}
 w_\alpha\propto {\cal L}_{\rm ChEFT}^{(\alpha)} \, {\cal
   L}_{M}^{(\alpha)} \, {\cal L}_{_{\rm GW}}^{(\alpha)} \, {\cal
   L}_{X}^{(\alpha)}\,, \qquad \sum_\alpha
 w_\alpha=1\,.\label{eq:S_weights}
\end{equation}
From the full ensemble we keep those EOSs that obey the Seidov condition,
that support stable stars with nonzero quark-core radius, and whose
metastable continuation reaches the overpressure needed for nucleation;
the resulting subset is visible in Fig.~\ref{fig:QMcoreEOSs}.

\section{PT description}
\label{sec:S2}

The PT is implemented through a Maxwell construction so that, at the
critical pressure $p_{\rm crit}$, the pressure and chemical potential are
continuous, whereas the number density and energy density exhibit finite
discontinuities. Consequently, the rest-mass density of such a stellar
configuration will have a sharp discontinuity at the quark-matter core
radius $R_q$, where the density will be double-valued, depending on
whether the interface between the two phases is approached from the
hadronic or the quark-matter side. Denoting with $\mathcal{Q}(r)$ the
local fraction of matter that is in the quark phase, which is different
from the fraction $q(t)$ in Eq.~\eqref{eq:q(t)} that measures instead the
amount of metastable hadronic matter inside the supercritical core, its
radial behaviour will be that of a step function
\begin{equation}
  \mathcal{Q}(r)=
  \begin{cases}
    1, & r < R_q\,,\\
    0, & r > R_q\,,
  \end{cases}
\end{equation}
where the value of $\mathcal{Q}$ at $r=R_q$ is just a matter of
convention and has no physical consequence. We note that the Maxwell
construction does not produce a finite-width equilibrium ``mixed phase''
where both phases coexist. Rather, after the PT, coexistence occurs only
at the codimension-one interface $r=R_q$, which has a zero-volume
measure. As a result, the equilibrium stellar solution contains hadronic
and quark-matter regions separated by a sharp phase boundary, rather than
an extended region with intermediate values $0 < \mathcal{Q} < 1$.

The metastable continuation of each EOS begins at the critical chemical
potential $\mu_{\rm crit}$, as shown in the inset of Fig.~\ref{fig:fig_1}
of the main text. We construct it by extrapolating the linear sound-speed
segment immediately below the PT to $\mu>\mu_{\rm crit}$ and
reconstructing the corresponding pressure, number density, and energy
density using Eqs.~\eqref{eq:S_nmu} and~\eqref{eq:S_pmu}. This
continuation does not represent the globally stable equilibrium EOS:
above $\mu_{\rm crit}$, the quark-matter branch is thermodynamically
preferred, while the extrapolated low-density branch is interpreted as
metastable. Note that the squared sound speed can have a relatively large
slope at the PT, and therefore eventually exceed the causal
limit, $c_s^2 \leq 1$, or reach the spinodal instability $c_s^2 \geq 0$. 
We terminate the metastable continuation at the
chemical potential $\mu_{\rm max}$ at which either limit is
reached, or at $500\,\mathrm{MeV}$ (whichever is first). Clearly, the 
continuation can also be terminated at a lower
chemical potential if the corresponding TOV solution becomes radially
unstable. Following this construction, for supercritical chemical
potentials $\mu_{\rm crit} < \mu < \mu_{\rm max}$ two distinct locally
stable branches are available, corresponding to the hadronic and
quark-matter phases. The overpressure is thus defined as $\Delta p(\mu)
:= p_q(\mu)-p_h(\mu)$, and is the driver of the nucleation process
described in Eqs.~\eqref{eq:rate}--\eqref{eq:q(t)} of the main text.

The ``compression history'', that is, the thermodynamic trajectory of
the stellar model from being entirely in the stable hadronic phase, over
to the metastable region, and then across the PT to the new phase, is
motivated in our analysis by the accretion of infalling matter onto the
PNS, and it is described by the mass-accretion history $M_b(t)$. This
quantity is well-measured in some numerical simulations, see
\eg~\cite{Shankar2025}, but the rate $\dot{M}_b(t)$ at which the star
accretes matter has a strong impact on the nucleation process, and its
universal behaviour is presently unknown. For these reasons, we model
the rest-mass growth as a phenomenological law until it reaches the
critical value $M_b(t_{\rm crit})$, and parameterise the accretion-driven
overcompression along the metastable branch as a linear function:
\begin{equation}
  M_b(t) =
\begin{cases}
M_{b,0} + 0.1285\, t^{0.3} + \dfrac{0.1}{1 + 200/t} \quad& t \le t_{\rm crit}, \\[2ex]
M_b(t_{\rm crit}) + \dfrac{1}{1000}\dot{M}_b(t - t_{\rm crit}) \quad& t > t_{\rm crit},
\end{cases}
\label{eq:accretion_history}
\end{equation}
where $t$ is measured in milliseconds, $M_{b,0}$ is fixed in our analysis
to $1.1M_\odot$ but can be easily varied, and we take $\dot{M}_b$ to be
in the range $\in [0.2, 1.0]\,M_{\odot}/{\rm s}$; the phenomenological
form, the initial mass, and the range for $\dot{M}_b$ are all in
agreement with the results of Ref.~\cite{Shankar2025}\footnote{Note that
there is a typo in the Table~2 of Ref.~\cite{Shankar2025}, where the
accretion rate is expressed in units of $M_{\odot}/{\rm ms}$.}. In
essence, as the mass accretion process takes place and the stellar model
follows its thermodynamic trajectory, we mark the critical time $t_{\rm
  crit}$ at which \hbox{$\Delta p(t_{\rm crit}) := p_q[\mu_c(t_{\rm
      crit})]-p_h[\mu_c(t_{\rm crit})] = 0$}. At this time the bubble
nucleation rate is zero since the nucleation action, which is inversely
proportional to the cube of the pressure difference $\Delta p$, diverges.
However, as time proceeds, the probability of nucleation, which depends
nonlinearly on the bubble surface tension $\sigma$, becomes nonzero and
keeps growing until, at a later time $t_{\rm nuc} > t_{\rm crit}$, the
quark fraction grows rapidly as described by Eq.~\eqref{eq:q(t)}. Note
that we define the nucleation time $t_{\rm nuc}$ as the time when the
metastable fraction of the star reaches $q(t=t_{\rm nuc})=1/2$. This
threshold is mostly arbitrary and different values could be chosen;
however, because the conversion is so rapid compared to the accretion
timescale~\cite{Blas:2022xco} and $q$ goes from one to zero in a
timescale $\ll\mathcal{O}(\rm{ms})$, they would not change our results in
any significant manner. Finally, we note that the conserved quantity
across the PT is the baryon mass, not the gravitational mass. Hence, the
pre- and post-conversion stellar configurations have the same $M_b$ and,
in the case of rotating configurations, the same angular momentum $J$.

\section{Gravitational-wave modelling}
\label{sec:S3}

We next illustrate in short how the multi-band GW emission is modelled in
the low- and high-frequency bands.

\subsection{kHz Emission}
\label{sec:S3.1}

We model the kHz component as the quadrupolar $\ell=2,\, m=0$ emission
from the $F$-mode (also referred to as
$^2f$-mode~\cite{Abdikamalov2009b}) oscillations following the rapid
structural readjustment of the stellar core after the PT. Using the
universal relations derived in Ref.~\cite{Zhao:2022tcw}, we can express
the frequency and damping time as
\begin{align}
 \fkHz&=\frac{1}{2\pi\,M}  \sum_{i=0}^{7}a_i(\ln\Lambda_q)^i\,,
 \label{eq:S_Ref}\\
 \tau_{_{F}}&= M \left[ \sum_{i=0}^{7}b_i(\ln\Lambda_q)^i \right]^{-1}\,,
 \label{eq:S_Imf}
\end{align}
where the coefficients $a_i$ and $b_i$ are tabulated in
Ref.~\cite{Zhao:2022tcw} and $\Lambda_q$ is the tidal deformability of
the hybrid star. The values of the damping time obtained in this way are
also consistent to a $5\%$ precision with those obtained for hadronic
EOSs using the phenomenological expressions presented in
Ref.~\cite{Lioutas:2017xtn}
\begin{equation}
 \frac{M}{\tau_{_{F}}}
 \simeq0.112\,\mathcal{C}^4-0.53\,\mathcal{C}^5+0.628\,\mathcal{C}^6,
 \label{eq:S_taufC}
\end{equation}
where $\mathcal{C}:=M/R$ is the stellar compactness. The dimensionless
quadrupole is obtained from the $I$--Love--$Q$ fit~\cite{Yagi2013a,
  Zhao:2022tcw} in terms of the tidal deformability $\Lambda$
\begin{equation}
 \ln{Q}(\Lambda)=\sum_{i=0}^{4}c_i(\ln\Lambda)^i \,,
 \label{eq:S_ILQ}
\end{equation}
from which we can compute the quadrupole difference $\Delta Q
:={Q}(\Lambda_h) - {Q}(\Lambda_q)$. We fix the dimensionless spin of the
star before the PT $\chi := J/M^2$ to a ballpark value of $\chi=0.2$,
although we note that the $I$--Love--$Q$ relations are also valid outside
the slow-rotation regime as long as the magnetic fields are
weak~\cite{Doneva2014a, Pappas2014, Haskell2014}.

For an equatorial observer, the measured GW signal in the kHz band
can then be simply expressed as a damped sinusoidal signal
\begin{equation}
 h_{\rm k} (t) =h_0 \, e^{-t/\tau_{_{F}}}\cos(2\pi\fkHz t)\,,
 \label{eq:ringdown}
\end{equation}
where the mode amplitude is estimated as
\begin{equation}
  h_{\rm 0} = \eta_f\frac{4\pi^2 \fkHz^2}{d} M^3 \chi^2\Delta Q\,,
 \label{eq:h0}
\end{equation}
with $d$ being the source distance and $\eta_f \leq 1$ the fraction of
the equilibrium quadrupole change deposited into the $F$-mode. Such an
efficiency can only be calibrated after performing nonlinear numerical
simulations, but we here assume it to be $\eta_f = 1$ so as to have a
clear upper-limit estimate rather than an uncertain model-dependent
prediction.

The kHz GW signal expressed by Eq.~\eqref{eq:ringdown} is sufficiently
simple that the one-sided power spectral density can be defined
as~\cite{Moore2014} \footnote{Hereafter, the index $h$ does not refer to
the hadron component but follows the standard notation for the strain.}
\begin{equation}
 S_{h, {\rm k}} (f) := 4 f\,\big|\widetilde{h}_{\rm k}(f)\big|^2\,,
 \label{eq:PSD_conventions}
\end{equation}
where $\widetilde{h}_{\rm k}(f)$ is the Fourier transform of~\eqref{eq:ringdown}.
The power spectral density can therefore be expressed analytically as
\begin{equation}
 S_{h, {\rm k}} (f)=\frac{f \, h_0^2 \, \tau_{_{F}}^2}
 {1+4\pi^2(f-\fkHz)^2 \,\tau_{_{F}}^2}\,,
 \label{eq:S_Shf}
\end{equation}
and it is such that the square of its \textit{characteristic strain},
which we define as
\begin{equation}
  h_{\rm ch}^2 := \int_0^\infty S_h(f) \, df\,,
  \label{eq:S_hcf}
\end{equation}
is, for the kHz signal, given by the simple algebraic
expression
\begin{equation}
  h_{\rm ch,k}^2\simeq\frac{1}{2} h_0^2 \, \fkHz \, \tau_{_{F}} \,.
\end{equation}
Similarly, it is possible to define a \textit{characteristic frequency}
as
\begin{equation}
  \ln f_{\rm ch} :=\frac{1}{h_{\rm ch}^2}\int_0^\infty \,\ln f\,S_h(f)
  \,df,
 \label{eq:S_fc}
\end{equation}
where, because of the simplicity of the corresponding GW signal, the kHz
characteristic frequency is given by $f_{\rm ch,k} \simeq f_{_{F}}$. As a
final remark, we note that since the characteristic strain $h_{\rm ch,k}
\sim h_0$, the kHz signal will depend on the EOS (via $Q$), on the
stellar angular momentum (via $\chi$), but also on $\sigma$, which fixes
how far beyond $M_{\rm crit}$ the star is driven before nucleation and
hence the value of the mass $M$ at the PT.

\subsection{MHz Emission}
\label{sec:S3.2}

For a transient source at distance $d$, and in analogy with
Eq.~\eqref{eq:PSD_conventions}, we define the one-sided power spectral
density of the MHz signal
\begin{equation}
 S_{h, {\rm M}} (f) := 4 f\,\big|\widetilde{h}_{\rm M}(f)\big|^2\,,
 \label{eq:PSD_conventions_M}
\end{equation}
where
\begin{align}
 \big|\widetilde{h}_{\rm M}(f)\big|^2 &:= \frac{R_q^2}{d^2}\frac{1}{\pi
   f^4} {\cal A}\,\widetilde P_{_{\rm GW}}(2\pi f R_{\rm bbl})\,,
 \label{eq:S_hMHz}
\end{align}
where ${\cal A} \, \widetilde{P}_{\rm GW}$ is the GW energy spectrum per
logarithmic frequency, and its amplitude and shape are defined as
\begin{align}
  {\cal A} & :=8\pi \rho_{\rm kin}^2 \, R_q \, R_{\rm bbl}^4 \, (4\pi f^3)\,,
  \label{eq:S_ampl}\\
  \widetilde{P}_{\rm GW}(s) & := \widetilde{\Omega}\,
  \left(\frac{s}{s_p}\right)^3\,
  \left(\frac{7}{4+3(s/s_p)^2}\right)^{7/2}\,,
  \label{eq:S_shape}
\end{align}
where $\widetilde{\Omega}$ is the dimensionless efficiency of fluid
motion converted to GWs, which we set to $\widetilde{\Omega} \simeq
0.012$ following simulations of the sound-shell
model~\cite{Hindmarsh2021}, and $s_p$ is a factor which sets the position
of the spectral peak and that is found to be of $\mathcal{O}(10)$ in
simulations~\cite{Hindmarsh:2017Shape}. We set it to $s_p=5$ as our
velocities are generally much smaller than the ones employed in previous
simulations. The quantity $\rho_{\rm kin}$ is the fluid kinetic-energy
density, which we set to be a fraction $\mathcal{S}^4_{\rm nuc}/10$ of
the nucleation scale, and which is consistent with an estimate of $v_f^2
(e+p)$ for fluid velocities of $v_f\sim 0.2$~\cite{Bleau:2026ala}. From
the spectrum of \eqref{eq:S_hMHz}, and using the definitions
\eqref{eq:S_hcf} and \eqref{eq:S_fc}, we obtain the characteristic strain
and frequency $\left(h_{\rm ch,M}, \, f_{\rm ch,M}\right)$.

While we have opted for a simple and approximate estimate of the MHz GW
signal, a more refined estimate is obviously possible, and it would
amount to replacing Eqs.~\eqref{eq:S_hMHz}--\eqref{eq:S_shape} with
spectral quantities computed from single-bubble hydrodynamic profiles
(see \eg \cite{Bea2024, Bleau:2026ala}) and include details about the
finite lifetime of the source, possibly incomplete conversion, and
deviations from linear acoustics. Also, the sound-shell model of which we
employed a simplified formula was developed for cosmological PTs, and is
used here as an effective description of the acoustic source. Its
standard regime of applicability is the one of a transition from high- to
low-density phase, and a more complete treatment needs to take into
account the differences between these setups~\cite{Barni2024,
  Bleau:2026ala}.

\end{document}